\documentclass[runningheads,a4paper]{llncs}

\usepackage{graphicx}
\usepackage[space]{grffile}
\usepackage{latexsym}
\usepackage{textcomp}
\usepackage{booktabs}
\usepackage{url}
\usepackage{hyperref}
\hypersetup{colorlinks=false,pdfborder={0 0 0}}
\newif\iflatexml\latexmlfalse

\usepackage[english]{babel}

\usepackage[utf8]{inputenc}

\begin{document}

\mainmatter  

\title{Requirements quality assurance in industry: why, what and how?}


%
%

\author{Michael Unterkalmsteiner%
\and Tony Gorschek}

\institute{Software Engineering Research Lab Sweden, Blekinge Institute of 
Technology, Karlskrona, Sweden\\ \email{mun|tgo@bth.se}
}

%
%

\toctitle{Lecture Notes in Computer Science}
\tocauthor{Requirements quality assurance: what can we automate and is it 
worthwhile?}
\maketitle

\begin{abstract}
[Context \& Motivation] Natural language is the most common form to specify 
requirements in industry. The quality of the specification depends on the 
capability of the writer to formulate requirements aimed at different 
stakeholders: they are an expression of the customer's needs that are used 
by analysts, designers and testers. Given this central role of requirements as 
a mean to communicate intention, assuring their quality is essential to reduce 
misunderstandings that lead to potential waste.
[Problem] Quality assurance of requirement specifications is largely a manual 
effort that requires expertise and domain knowledge. However, this demanding 
cognitive process is also congested by trivial quality issues that should not 
occur in the first place.
[Principal ideas] We propose a taxonomy of requirements quality assurance 
complexity that characterizes cognitive load of verifying a quality aspect from 
the human perspective, and automation complexity and accuracy from the machine 
perspective. [Contribution] Once this taxonomy is realized and validated, it 
can serve as the basis for a decision framework of automated requirements 
quality assurance support.
\end{abstract}%


\section{Introduction}
The requirements engineering process and the artefacts used in 
coordination and communication activities influence the performance of 
downstream development 
activities~\cite{damian_empirical_2006}. While 
research has proposed myriads of formal, semi-formal and informal methods to 
convey requirements, plain natural language (NL) 
is the \emph{lingua franca} for specifying requirements in 
industry~\cite{sikora_industry_2012,kassab_state_2014}. One potential 
reason is that NL specifications are easy to 
comprehend without particular training~\cite{carew_empirical_2005}.
However, NL is also inherently imprecise and ambiguous, posing challenges in 
objectively validating that requirements expressed in NL represent the 
customers' needs~\cite{ambriola_processing_1997}. Therefore it is common 
practice to perform some sort of review or inspection~\cite{kassab_state_2014} 
to quality assure NL requirements specifications.
While there exists a plethora of methods to improve requirements 
specifications~\cite{pekar_improvement_2014}, there are no guidelines that 
would support practitioners in deciding which method(s) to 
adopt for their particular need. We think that a first step to such a decision 
framework is to characterize the means by which quality attributes in 
requirements specifications can be affected. Therefore, we initiated an applied 
research collaboration with the Swedish Transport Administration (STA), the 
government agency responsible for the rail, road, shipping and aviation 
infrastructure in Sweden. STA's overall goal is to improve the 
communication and coordination with their suppliers, mostly handled through NL 
requirements specifications. Infrastructure projects vary in duration (months 
to decades) and budget (up to 4 Billion USD), requiring an adaptive quality 
assurance strategy that is backed by methods adapted to the needs of the 
particular project. The large number of requirements (several thousands) and 
the need to communicate them to various suppliers makes specifications in NL 
the only viable choice. Still, STA needs to quality assure the requirements and 
decide what level of quality is acceptable.
In this paper we present the basic components for a taxonomy that will drive, 
once the research is completed, 
a requirements quality assurance decision support framework. To this end, we illustrate a research outline aimed at answering our 
overall research question: \textbf{How can we support practitioners in achieving ``good-enough'' requirements specification quality?}


\section{Related Work}\label{sec:relatedwork}
Davis et al.~\cite{davis_identifying_1993} proposed a comprehensive set of 24 
attributes that contribute to software requirements specification (SRS) 
quality. Saavedra et al.~\cite{saavedra_software_2013} compared this set with 
later contributions that studied means to evaluate these attributes. 
Similarly, Pekar et al.~\cite{pekar_improvement_2014} reviewed the literature 
and identified 36 studies proposing techniques to improve SRS quality. While 
Agile software development is notorious for promoting as little documentation 
as possible~\cite{fowler_agile_2001}, Heck and 
Zaidman~\cite{heck_systematic_2016} identified 28 quality criteria used for 
Agile requirements, six of them being novel and specifically defined for Agile 
requirements. All these reviews point to relevant related work potentially 
contributing to the components of a decision support framework for requirements 
quality assurance. The importance of providing decision support to 
practitioners is growing hand-in-hand with the complexity of today's developed 
software products and the available number of technologies to realize 
them~\cite{hassan_roundtable:_2013}. To the best of our knowledge, no framework 
exists to support the selection of requirements quality assurance techniques.

\section{Characterizing Requirements Quality Assurance}\label{sec:taxonomy}
\begin{figure}
\begin{center}
\includegraphics[width=0.5599999999999999\columnwidth]{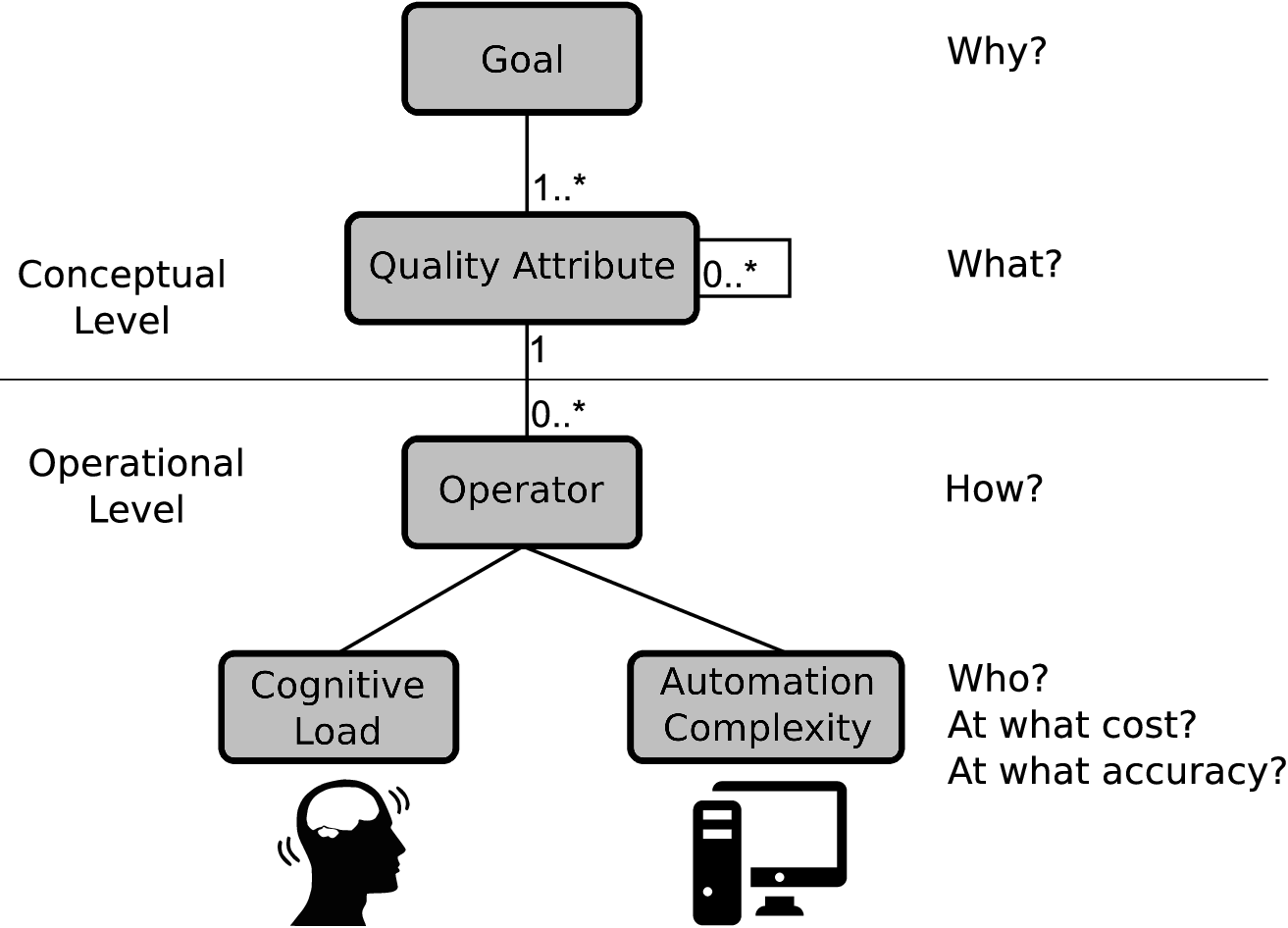}
\caption{{\label{fig:tax} Requirements quality assurance taxonomy%
}}
\end{center}
\end{figure}

The purpose of this taxonomy is to characterize the 
components that are involved in the process to achieve a particular 
requirements quality (RQ) level (Figure~\ref{fig:tax}). This systematization 
then allows to take informed decisions about effort and potential impact for 
RQ improvement. 

A \emph{goal} determines what the improvement of RQ should achieve. Typical 
goals could be to improve the communication between stakeholders, to improve 
the ability to verify the product, or better cost estimates. Different goals 
can also contradict each other. Goals are important as they provide a scope 
that limits the potential actions on the operational level to a set that is 
economically acceptable - this enables focus of efforts to assure certain 
quality aspects within the given opportunities of the resources afforded.  

\begin{figure}
	\begin{center}
		\includegraphics[width=0.98\columnwidth]{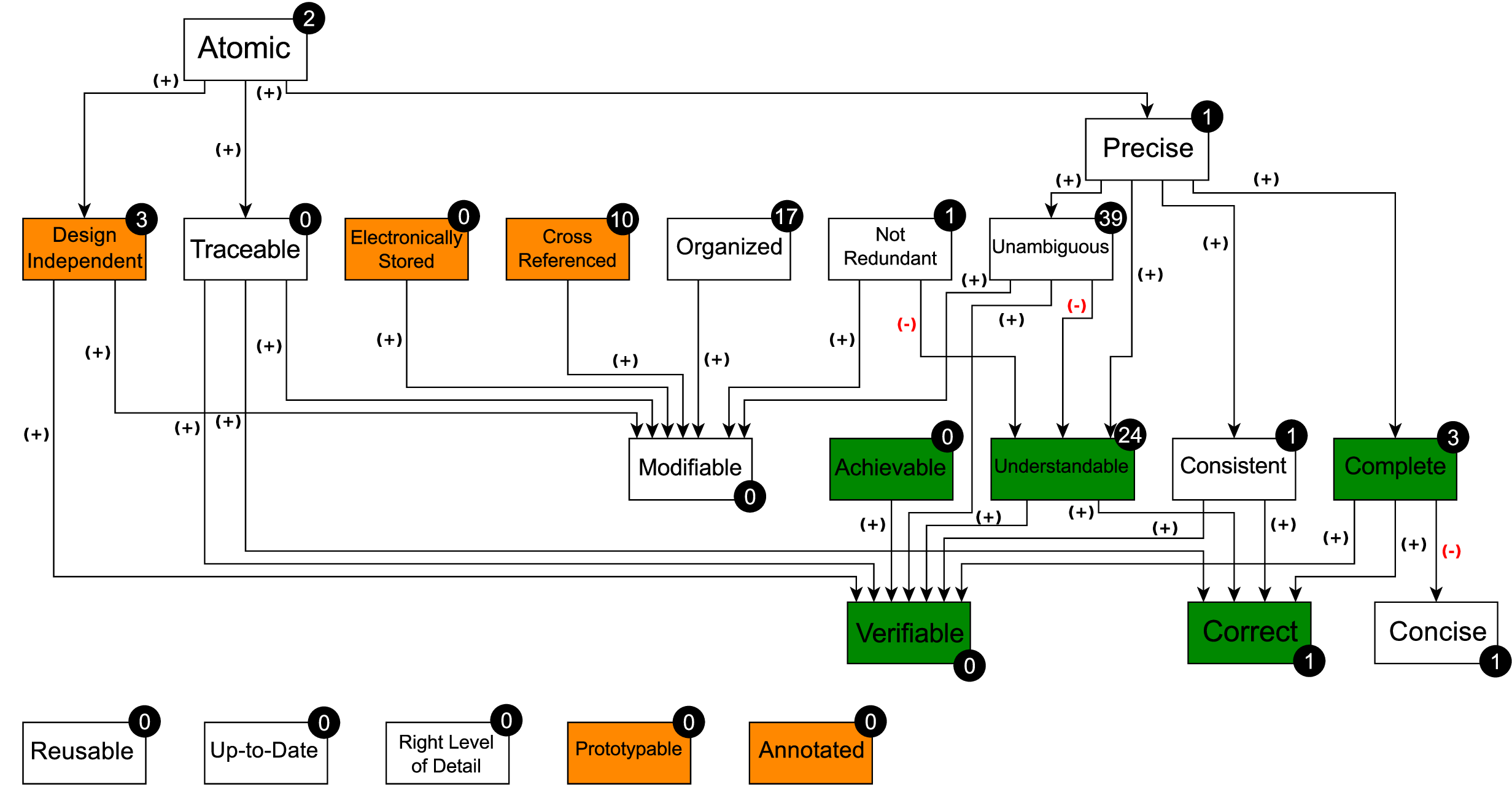}
		\caption{{\label{fig:qa} Quality attributes and their relationships 
				(adapted 
				from Saavedra et al.~\cite{saavedra_software_2013}; color 
				coding and numbers 
				are our addition, and used and explained in 
				Section~\ref{sec:outline})%
		}}
	\end{center}
\end{figure}

\emph{Quality attributes} describe the favourable properties of a 
requirement. For example, unambiguity is commonly defined as the quality of a 
statement being interpretable in a unique way. 
Quality attributes for requirements have been described in numerous quality 
models, reviewed by Saavedra et al.~\cite{saavedra_software_2013}. Quality attributes are not independent, 
i.e. one attribute can positively or negatively influence another. 
Figure~\ref{fig:qa} provides an overview of RQ attributes and 
their relationships to each other. For example, atomicity positively 
influences design independence, traceability and precision of a requirement, as 
indicated by the (+) in Figure~\ref{fig:qa}. On the other hand, unambiguous 
requirements, often achieved by higher formality, are generally also less 
understandable. 

Goals and quality attributes build the \emph{conceptual level} of the taxonomy. 
They can help to answer questions pertaining to why an improvement of 
RQ is necessary, and what quality attributes are associated with that goal. 
Taking the example from earlier, improving the ability to verify the 
product based on the stated requirements, one can see in Figure~\ref{fig:qa} 
that many quality attributes influence requirements verifiability. Depending on 
constraints in the operational level, discussed next, one can decide how to 
reach the stated goal by choosing a set of quality attributes, which in turn 
are associated with operators.

\emph{Operator} is the generic term we use for instruments that tangibly
characterize quality attributes. An operator provides a definition of how a 
requirement is analysed w.r.t. the associated quality attribute. Examples of 
operators are metrics~\cite{fabbrini_automatic_2001,genova_framework_2011}, 
requirement smells~\cite{femmer_rapid_2016} or rules and constraints on how to 
formulate requirements. An operator can be implemented by either a person or a 
computer program (or both). In either case, we want to characterize the 
operator by some notion of cost and accuracy, providing input for the 
decision on how and whether at all to realize the operator.
We borrow the concept of \emph{cognitive load} from the field of 
instruction design where cognitive load theory~\cite{sweller_intrinsic_2011} is 
used to describe and improve learning efficiency. Each operator is associated 
with a level of intrinsic cognitive load, describing the complexity of applying 
the operator on a single requirement or a complete specification. For example, 
if the operator is the ambiguous adverbs requirements 
smell~\cite{femmer_rapid_2016}, then the intrinsic cognitive load is determined 
by the number of ambiguous terms one has to remember to detect these terms in 
the requirements text. Since cognitive load is 
additive~\cite{sweller_intrinsic_2011}, there are 
(individual) limits to the efficiency of applying operators, and is therefore 
one determinant for the effective cost of RQ assurance.
If an operator is realized through machine-based processing of information, 
we characterize this realization by its \emph{automation complexity}. 
Continuing with the example of ambiguous adverbs, the 
automation complexity of this operator is low as it can be implemented with a 
dictionary~\cite{femmer_rapid_2016}. On the other 
hand, some of the requirements writing rules found in STA are rather complex. 
For example, one rule states that repetition of requirements shall be avoided 
and a reference to a general requirement shall be made (addressing redundancy). 
The detection of rule violations requires the analysis of the complete 
specification, identifying similar phrased statements. While 
this is certainly possible (e.g. with code clone and plagiarism 
detection~\cite{chen_shared_2004}), the analytical complexity is 
higher than for a dictionary lookup.

\section{Research Outline}\label{sec:outline}
The taxonomy serves three main purposes which are outlined in this section, 
together with six research questions and our planned approaches to answer them.

\subsection{Prioritize quality attributes}
We have asked six requirements experts at STA to rank RQ 
attributes (definitions were extracted from the review by Saavedra et 
al.~\cite{saavedra_software_2013}) by their importance 
using cumulative voting~\cite{berander_requirements_2005}. Figure~\ref{fig:qa} 
shows the five top and bottom attributes in green and orange respectively. 
Individual quality attributes have been researched earlier, focusing on 
ambiguity, completeness, consistency and 
correctness~\cite{pekar_improvement_2014}. While the 
perceived importance of completeness and correctness is matched by research on 
these attributes, ambiguity and consistency were ranked by the experts only at 
position 13 and 16 respectively. At first sight, this might indicate that 
research focus needs adjustment. However, taking into consideration the 
relationships between quality attributes, we see a moderate overlap between the 
needs at STA and past research. Nevertheless, there are certain 
quality attributes whose evaluation has seen little research, like 
traceability~\cite{pekar_improvement_2014}, while being important for STA since 
they affect verifiability and correctness. The relationships between quality 
attributes inform us also about potential inconsistencies among the goals 
of quality improvement. For example, design independence was ranked by STA's 
experts on position 21 while it affects verifiability, 
ranked at position 3. This could indicate that, while verifiability is 
important for STA, design independence as a related aspect has been overlooked as a means to achieve this.
These examples show how the relationships between quality attributes can be 
used to analyse the goals of the company. However, since Saavedra et 
al.~\cite{saavedra_software_2013} deduced the relationships shown in 
Figure~\ref{fig:qa} by interpreting the quality models they reviewed,
these dependencies need further empirical validation, leading to \emph{RQ1: To 
what extent do requirements quality attributes affect each other?} One approach 
to address this question, dependent on the answers to the questions in 
Section~\ref{sec:ops}, would be to analyse 
the correlation between operators for different quality attributes. We plan to 
perform this analysis at STA, which in turn partially answers \emph{RQ2: To 
what extent can quality attribute rankings be used for planning quality 
assurance activities?} Further inquiries at STA are needed to identify factors 
that affect planning, such as timing (does quality attribute importance depend 
on the project phase?) and implementation cost.

\subsection{Determine operators and their accuracy}\label{sec:ops}
At STA we have identified 110 \emph{operators} in the form of requirements 
writing rules. These rules describe how requirements shall 
be formulated and provide review guidelines. 
Table~\ref{tab:rwr} shows five examples of writing rules. We have mapped, 
where the description allowed it, which quality attribute was primarily 
targeted by each rule. The numbers in Figure~\ref{fig:qa} indicate 
how many operators we identified for each quality attribute. Several quality 
attributes have no or very few associated operators, leading to the question 
\emph{RQ3: Which quality attributes can be characterized by an operator?} We 
plan to answer this question by systematically reviewing the literature, 
extending the work by Saavedra et al.~\cite{saavedra_software_2013}, 
Pekar et al.~\cite{pekar_improvement_2014}, and Heck and 
Zaidman~\cite{heck_systematic_2016}. On the other hand, we have identified 110 
operators in STA, leading to the questions \emph{RQ4: How can NL processing be 
used to implement operators?} and \emph{RQ5: What is the accuracy of these 
operators in relation to state-of-practice?} We estimated that  
40-50\% of the writing rules in STA can be implemented with current techniques, 
e.g. as proposed by Femmer et al.~\cite{femmer_rapid_2016}. However, as 
indicated in the last column of Table~\ref{tab:rwr}, techniques to implement
rules 4 and 5 still need to be determined. In addition we plan to evaluate the 
practical benefits of machine-supported RQ assurance compared 
to the state-of-practice, i.e. manual quality assurance, at STA.
\vspace{-5mm}
\begin{table}
	\scriptsize
	\caption{Examples of requirements writing rules at STA}\label{tab:rwr}
	\begin{tabular}{p{77mm}lp{20mm}}
		\toprule
		Rule & Quality Attribute & Implementation\\
		\midrule
		1. No time should be specified in the technical documents. Instead, 
		refer to the {Schedule} document. & Non-redundant & Named entity 
		extraction\\
		2. Numbering of figures, illustrations and tables should be 
		consecutively numbered throughout the document, starting from 1. & 
		Organized & Document meta-data analysis\\
		3. Numbers ``1-12'' shall be written as shown in the following example, 
		``to be at least two (2).'' & Unambiguous & POS Tagging\\
		4. Terms such as ``user'', ``dispatcher'', ``operator'' should be used 
		consistently. & Unambiguous & TBD\\
		5. If a functional requirement is supplemented by additional 
		requirements to clarify fulfilment, these must be written as separate 
		requirements. & Atomic & TBD\\
		\bottomrule
	\end{tabular}
\end{table}
\vspace{-10mm}

\subsection{Estimate cognitive load and automation complexity}
Applying all 110 operators on a specification consisting of thousands of 
requirements is a cognitively demanding task. For deciding how to implement an 
operator, it would be useful to be able to estimate the cognitive load each 
operator will cause and the complexity to implement the operator in a 
computer-based support system, leading to \emph{RQ6: How can the cognitive load 
and automation complexity of an operator be estimated?}
Cognitive load could be approximated by a heuristic that 
describes whether the application of the operator requires domain knowledge or 
not, and to what extent context needs to be considered. Context 
could be defined as ``local'', referring to a single requirement, ``regional'' 
referring to a section or chapter in the specification, or ``global'' the whole 
specification and beyond, e.g. regulations and standards. There exist also 
multiple approaches to measure cognitive load directly~\cite{chen_robust_2016}.
Automation complexity could be estimated by categorizing operators on the 
linguistic aspect they address. Operators that require semantic understanding 
are more complex than operators that require syntactic or lexical analyses of a 
requirement. The least complex operators are statistical, i.e. analyses that 
work with letter, word or sentence counts. Since, to the best of our knowledge, 
no such characterization of operators exists, we plan to collaborate with 
experts from both neuropsychology and linguistics to perform literature reviews 
and design experiments.

\section{Conclusion}\label{sec:conclusions}
In this paper, we have proposed a requirements quality assurance taxonomy that, 
once the stated research questions are answered, forms the engine for a 
decision framework that allows companies to initiate or improve their 
requirements quality assurance program through (a) realizing the consequences 
of dependencies between quality attributes in their current manual activities 
for quality assurance, (b) mapping cognitive load to the prioritized actions 
for quality assurance, and (c) enabling the decision on the trade-off between 
manual and machine-supported quality assurance, given cost and accuracy of the 
choices. 

\bibliographystyle{splncs03}
\bibliography{bibliography/converted_to_latex}

\end{document}